\documentclass[fleqn,10pt]{arXiv}
\pdfoutput=1

\usepackage[english]{babel}
\usepackage{natbib}
\usepackage{float}
\usepackage{textcomp}

\definecolor{color1}{RGB}{120,0,0}
\definecolor{color2}{RGB}{0,20,20}
\definecolor{color3}{RGB}{0,0,0}

\usepackage{hyperref}
\hypersetup{hidelinks,colorlinks,breaklinks=true,urlcolor=color2,citecolor=color1,
            linkcolor=color1,bookmarksopen=false,pdftitle={Title},pdfauthor={Author}}

\JournalInfo{arXiv preprint}
\Archive{}

\PaperTitle{Trivial pursuits \\ \vspace{0.25cm} \normalsize{Comment on ``Nonlinear averaging of thermal experience predicts population growth rates in a thermally variable environment'' by Bernhardt et al.}}
\PaperSubtitle{Trivial pursuits}

\Authors{Tarik C. Gouhier\textsuperscript{1,*} and Pradeep Pillai\textsuperscript{1}}
\affiliation{\textsuperscript{1}\textit{Marine Science Center, Northeastern University, 430 Nahant Road, Nahant, MA 01908}}
\affiliation{*\textbf{Corresponding author}: tarik.gouhier@gmail.com}
\newcommand{\keywordname}{Keywords}

\Abstract{We demonstrate that the conclusions drawn by \citet{bernhardt2018a} regarding the
ability of nonlinear averaging to accurately predict organismal performance under
fluctuating temperatures are flawed because of a series of experimental and statistical
issues that include the presence of a hidden treatment effect, the use of a single
low frequency temperature fluctuation that could easily be tracked by the fast growing organism,
and the decision to quantify performance via population growth rate, a metric that can mask
significant variation in population size.}

\begin{document}
\flushbottom
\maketitle
\thispagestyle{firstpage}


\section*{Introduction}

\citet{bernhardt2018a} sought to determine whether historical effects associated with the temporal sequence of temperatures needed to be accounted
for when predicting population growth under fluctuating temperatures. To do so, the authors asked the following questions: (i) Does the `fallacy of
the averages' apply to algae experiencing fluctuating temperatures due to the nonlinear relationship between growth and temperature? (ii) Does
accounting for this nonlinear relationship but not historical effects allow algal growth under fluctuating temperatures to be predicted from their
growth under the relevant constant temperatures? (iii) How does temperature variability affect algal growth in populations around the globe?

To answer these questions, the authors devised an experiment whereby they exposed an algal species to a series of constant temperature regimes to
determine how its growth varied as a function of temperature. They then conducted a second experiment where they exposed the same algal species to a
series of fluctuating temperature regimes, each consisting of alternating, equal-length periods of low and high temperatures. Using this experimental
data, the authors showed that the growth of algae exposed to fluctuating temperatures could not be predicted by their growth at the corresponding
constant average temperature (the ‘fallacy of the averages’). However, growth under fluctuating temperatures could be predicted by taking the
(nonlinear) average of the growth observed under the relevant constant temperatures. The authors then showed how the ‘fallacy of the averages’ could
lead to biases when estimating algal growth from constant temperatures for different species around the globe. Although this paper superficially
checks all the right boxes (i.e., it ostensibly combines experiments and observations to test theory), it suffers from several fundamental issues
outlined below.

\section*{Experimental and conceptual flaws}
First and foremost, the experiment used by Bernhardt et al. was poorly designed because it was unlikely to detect the historical
effects of temperature and thus unable to provide a rigorous test of the central premise of their paper. This is because the
authors only tested the effects of a single type of temperature fluctuation: a square waveform whose 1-day period
corresponds to the generation time of the algae \citep{pena2005,bernhardt2018a}. Hence, the fast-growing algae were very likely
able to track the relatively slow temperature fluctuations, thereby preventing the emergence of any potential historical effects.

\begin{figure*}[!htb]\centering
\includegraphics[width=\linewidth]{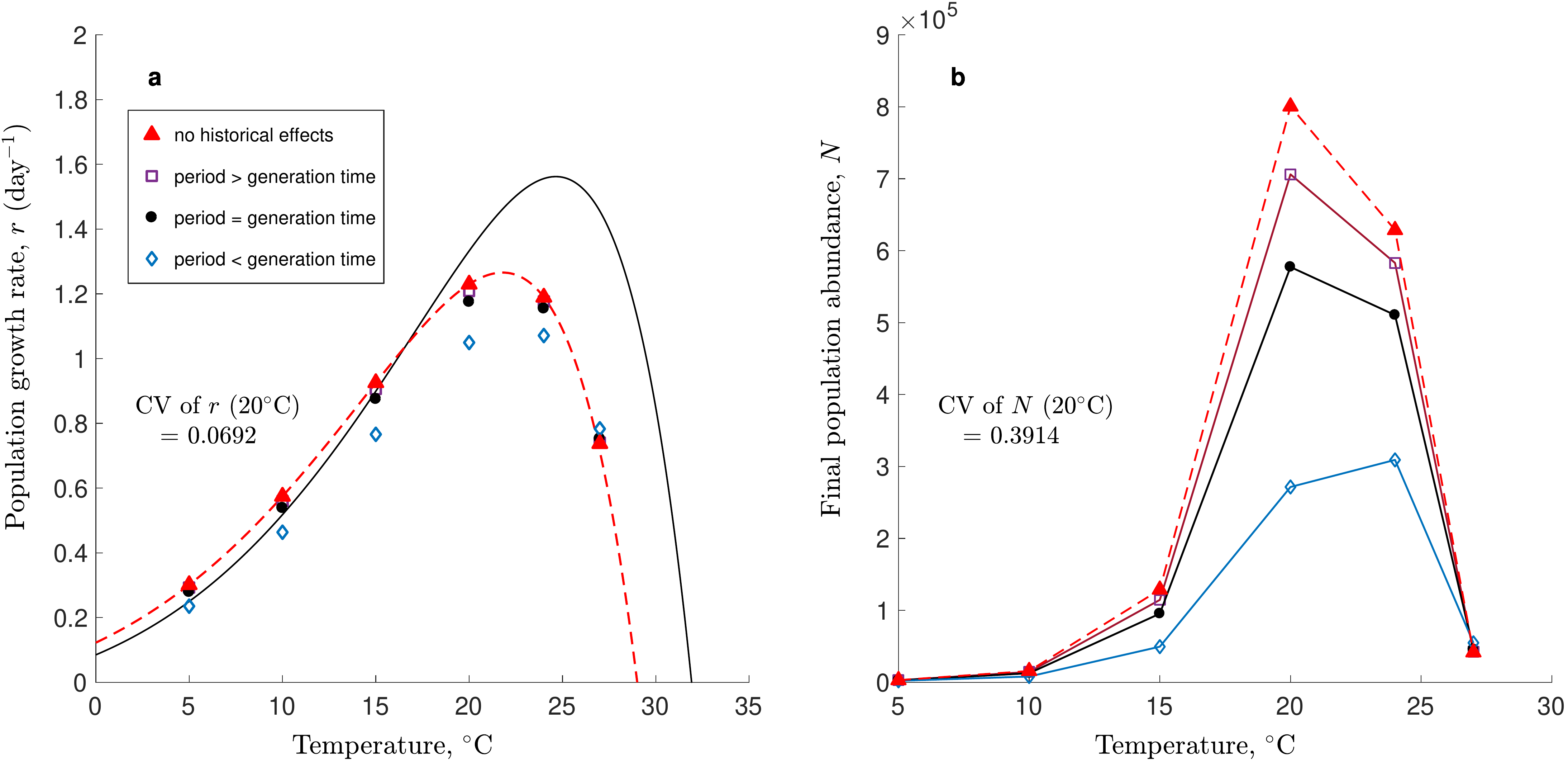}
\caption{Simulation of a stage-structured version of the continuous-time model presented in \citet{kremer2018} with an initial abundance of 500
individuals and an experiment duration of 6 days. (a) Thermal Performance Curves based on population growth rate $r$ are shown for constant
temperatures (solid line) and fluctuating temperatures based on nonlinear averaging (dashed red line). Simulations were run for temperature
fluctuations whose period was greater than (open squares), equal to (filled circles), or smaller than (open diamonds) the generation time of the
organism. Filled triangles falling on the dashed line (nonlinear averaging curve) are for simulations with no historical effects. The acclimatization
rate was set to $\sigma=10$ (simulations with a slower acclimatization rate of $\sigma=0.8$ were qualitatively identical, with the points being
slightly more spread-out vertically). (b) Final abundances from the same simulations show the extreme variation in $N$ (orders of magnitude).
Coefficients of variation (CV) for $r$ and $N$ across all temperature fluctuation periods at 20\textdegree C shown in (a) and (b).}
\label{fig1}
\end{figure*}

To illustrate this issue, we used simulations of a stage-structured version of the continuous-time model presented in \citet{kremer2018} under a
fluctuating temperature regime whose period was either smaller than, equal to or greater than the generation time of an organism that partially
tracked temperature over time (i.e., had the potential to exhibit historical effects because of its relatively slow response to temperature
variation). The growth rate of the organism, $r$, at each temperature as well as both the amplitude and the mean of the temperature fluctuations used
in the simulations were identical to those in Bernhardt et al. The simulations show a consistent and unsurprising pattern: when the period of the
fluctuations is larger than or equal to the generation time, (nonlinear) averaging the growth rate observed under the relevant constant
temperature regimes is more likely to accurately predict the growth rate observed under fluctuating temperatures because the organism is
able to track the temperature variation (Fig. \ref{fig1}a). However, variation in $r$ exists, and reductions in the period of the temperature
fluctuations leads to larger discrepancies between the observed population growth rate and that expected based on nonlinear averaging.
Here, reducing the period of the temperature fluctuations decreases the accuracy of nonlinear averaging because the organism is unable to
keep up with the increasingly rapid pace of the temperature changes (Fig. \ref{fig1}a). These simulations thus show that historical
effects are less likely to emerge when the period of the fluctuations is larger than or equal to the generation
time because the organism is able to track such relatively slow temperature variation. Hence, by experimentally choosing conditions
that made the emergence of historical effects less likely, Bernhardt et al. were unable to provide a robust test of the hypothesis
that the temporal sequence of temperature fluctuations could be safely ignored when predicting population growth under fluctuating
temperatures via nonlinear averaging.

Even if historical effects could emerge despite the use of a single long-period fluctuation, additional issues in both the experimental
design and the statistical properties of the metric used to assess organismal performance would have made their detection unlikely.
Indeed, Bernhardt et al. grew their algae at 16\textdegree C for one year and then conducted both their 7-day constant and variable
temperature experiments without acclimatizing their organism, a procedure that is not standard
even in studies focusing on quantifying the effects of acute temperature change \citep{kremer2018}. This means that their constant temperature regime
was not actually constant. Instead, it represented a single ‘asymmetrical’ temperature fluctuation with a very large period (365 days at
16\textdegree C followed by 7 days at a new constant temperature). Additionally, their variable temperature regime consisted of two different
temperature fluctuations with distinct periods: the same ‘asymmetrical’ large period temperature fluctuation that was present in the constant
treatment (365 days at 16\textdegree C followed by 7 days of fluctuations around a new average temperature) and a small 1-day period fluctuation.
Hence, it is likely that the fluctuation with the large period, which was a hidden treatment common to both the constant and the variable temperature
regimes, masked the effects of the small period fluctuations, thus making it more likely that (nonlinear) averaging growth under constant
temperatures would accurately predict growth under variable temperatures. Finally, the use of the average population growth rate $r$ as a metric for
determining the accuracy of nonlinear averaging under variable temperatures is problematic because small differences in these rates can mask
extremely large differences in population densities between the constant and variable temperature regimes (Fig. \ref{fig1}b).

Hence, although Bernhardt et al. sought to explicitly test whether historical effects could be ignored when
predicting population growth rate under variable temperature regimes via nonlinear averaging, the suite of experimental and statistical
issues outlined above led them instead to implicitly test a set of trivial hypotheses. Specifically, by preventing the emergence of historical effects,
Bernhardt et al.’s experiment ended up being reduced to a test of the ‘fallacy of the averages’ and the accuracy of ‘nonlinear averaging’,
which are not testable hypotheses but mathematical inevitabilities associated with the properties of the arithmetic mean. Indeed, the
‘fallacy of the averages’ is a direct consequence of Jensen’s inequality \citep{jensen1906} which states that if $f(x)$ is a nonlinear
function of $x$, then the function of the average $f(\overline{x})$ is not equal to the average of the function $\overline{f(x)}$. This is
because the additivity and homogeneity properties defining linear functions do not hold for nonlinear functions, so the order of the
operations matters: averaging and then taking the function is not the same as taking the function and then averaging. Hence, the only way for
the ‘fallacy of the averages’ not to play-out in these experiments is for algal growth $f(x)$ to be a linear function of temperature $x$. In
that case, linearity would ensure that the order of the operations does not matter so the function of the average $f(\overline{x})$ would be
equal to the average of the function $\overline{f(x)}$ However, decades of research on thermal performance across a multitude of organisms
and environments demonstrates that the relationship between growth and temperature is almost universally nonlinear (specifically unimodal and
asymmetrical). Hence, contrary to the authors’ claims, the ‘fallacy of the averages’ does not represent a hypothesis that needs to be tested
but an inevitable result that was already well established and whose implications for the field of ecology were reviewed over twenty years
ago \citep{ruel1999}.

The second hypothesis, which posits that nonlinear averaging accurately predicts growth under fluctuating temperatures, is equally flawed.
This is because the ‘fallacy of the averages’ and ‘nonlinear averaging’ are not independent hypotheses but complements. Hence, if not
accounting for the nonlinear relationship between growth and temperature ensures inaccurate predictions due to the ‘fallacy of the averages’,
then accounting for it via nonlinear averaging guarantees accurate predictions in the absence of historical effects. Indeed, nonlinear
averaging amounts to nothing more than taking the arithmetic mean of a nonlinear function, and the arithmetic mean applies equally well to
linear and nonlinear functions. In this case, if an organism’s growth is $f(x_1)$ under constant temperature $x_1$ and $f(x_2)$ under
constant temperature $x_2$, then its average growth under a variable temperature regime consisting of two time periods of equal length
characterized by temperatures $x_1$ and $x_2$, respectively, will simply be the arithmetic mean $\frac{f(x_1) + f(x_2)}{2}$. This will be
true regardless of the nonlinearity of function $f(x)$ with respect to temperature $x$. Hence, in the absence of historical effects, there
was never any doubt that (nonlinear) averaging algal growth under the relevant constant temperature regimes would accurately predict algal
growth under fluctuating temperatures. Framing the results of the experiment in terms of Jensen’s inequality simply served to obscure their
obvious and trivial nature. Overall, this entire exercise boils down to (1) selecting a biological function that is known to be nonlinear so
that the order of the operations matters, (2) performing the operations in the wrong order and thus getting the wrong results (‘fallacy of
the averages’), and then (3) performing the operations in the right order and thus getting the right results (‘nonlinear averaging’).

The only way that the outcome of the experiment could have been interesting is if the predicted mean growth under variable temperatures were
not equal to the arithmetic mean of the growth observed under the relevant constant temperature regimes. This could happen if temperature
fluctuations had sufficiently large historical effects so as to shift the growth curve from $f(x)$ under constant temperatures to $g(x)$
under variable temperatures. Here, the predictions could fail because one would be applying the arithmetic mean to function $f(x)$ in order
to estimate the mean of function $g(x)$. However, the authors ensured that this would not be likely by selecting an organism with a high
growth rate and a short generation time relative to the period of the temperature fluctuations. This and other experimental decisions ensured
the triviality of their results.

Aside from the inevitability of the core hypotheses that ended up being tested, the paper also suffers from a few additional issues. For
instance, the authors showed that a different approach based on what they called scale transition theory--—but is actually just Taylor
expansion—--produced accurate estimates of growth under variable temperatures when the mean and the variance of temperature were available
but not its underlying time series. However, in what situation are the mean and the variance of temperature available, but not the underlying
time series that they are based on? This seems like an unlikely and contrived scenario meant to obfuscate the obviousness of the main result
(i.e., that one can take the arithmetic mean of a nonlinear function). Additionally, if the goal is to make accurate predictions, why not
include higher order terms in the Taylor expansion? The only justification seems to be to accommodate an arbitrary constraint regarding data
availability.

The final issue with the paper is that the authors used nonlinear averaging to show how variable temperatures are likely to affect growth in
several algal species around the globe. However, these results are based on the assumption that temperature is the sole driver of growth.
Their approach does not account for differences in food availability or other factors such as light limitation that are likely to affect
growth. These results thus constitute a trivial ‘proof’ that nonlinear averaging can predict growth under the assumptions that (1)
temperature is the rate limiting step with respect to growth and (2) that historical effects are completely absent. Although the authors
acknowledged some of these issues and thus referred to their results as ‘first-order predictions’, they failed to recognize the extent and
severity of the limitations inherent in their approach. At best, their results provide unnecessary ‘strategic’ evidence for the ‘fallacy of
the averages’ and the accuracy of nonlinear averaging, both of which are well established and widely understood mathematical laws. At worst,
they provide inaccurate and downright misleading ‘tactical’ predictions about the particularities of algal growth that fail to account for
other biotic and abiotic factors. In other words, these predictions are not even useful as null hypotheses.

Overall, this paper represents an atypical use of mathematics in the natural sciences. In general, it is perfectly valid to conduct
experiments in order to test model predictions because the latter make simplifying assumptions about the natural world that can lead to
systematic discrepancies between theory and reality. However, Jensen’s inequality and nonlinear averaging are not theoretical models or
hypotheses, but simple mathematical properties of the arithmetic mean whose inescapability make them untestable in the classical sense. For
hypotheses to be useful, they must have a non-zero probability of being false. Otherwise, such hypotheses are merely trivial inevitabilities
masquerading as scientific uncertainties.

\phantomsection
\section*{Acknowledgments}

We acknowledge support from the National Science Foundation (OCE-1458150, OCE-1635989, CCF‐1442728).

\phantomsection
\bibliographystyle{ecology}
\bibliography{preprints}


\end{document}